\documentclass[12pt]{article}
\begin{document}

%УДК 539.12.01, $\quad$ PACS 12.15-y

\begin{center}
{\large \bf  The electroweak model with  rarely  interacted neutrinos}
\end{center}

\begin{center}
N.A.~Gromov \\
Department of Mathematics, Komi Science Center UrD, RAS, \\
Kommunisticheskaya st. 24, Syktyvkar 167982, Russia \\
E-mail: gromov@dm.komisc.ru
\end{center}

%\begin{center}
%Abstract
%\end{center}

The electroweak model, which  lepton sector correspond to 
%is characterized by 
the contracted gauge group 
$SU(2;j)\times U(1), \; j\rightarrow 0$,  whereas
boson and quark sectors are standard one, is suggested.
This model describe in a natural manner 
why neutrinos so rarely  interact  with matter, as well as why neutrinos 
cross-section increase with the energy.
Dimensionfull  parameter of the model is interpreted as neutrino energy.
Dimensionless contraction parameter $j$ for low energy is connected with the  Fermi constant of weak interactions and is approximated as $j^2\approx 10^{-5}$.

\vspace{5mm}

{\bf Keywords}: gauge theory; electroweak model; contraction; neutrino

PACS number:  12.15-y 

\section{Introduction }

The standard electroweak model  based on gauge group $ SU(2)\times U(1)$ gives a good  description of electroweak processes.
Due to this model  the W- and Z-bosons was predicted and experimentally observed at the end of the last sentury. Higgs boson is now searched at the modern  Large Hadron Collider.
At the same time the grave disadvantage of the standard model is the
 presence  about fifteen free parameters. 
But among these %does not present 
there is not such parameter, which a priori can be regarded as a small one and can be connected with the experimentally observed very rare interaction neutrinos with matter.

The purpose of this paper is to build up %suggest 
the variant of the electroweak model, which naturally describe the vanishingly small, as compared to other particles, interaction neutrinos with anything with the help of the zero tending contraction parameter, as well as to give the physical interpretation of the contraction procedure.
The model is suggested where the boson and quark sectors are the same as in the standard electroweak model, but the lepton sector %is invariant with respect to
 correspond to the  gauge group $SU(2;j)\times U(1)$, where $j\rightarrow 0$ is dimensionless contraction parameter.

\section{Standard electroweak model}

The Lagrangian of the standard electroweak model is given by the sum 
$$ %\begin{equation}
L=L_B  + L_Q + L_L %=L_A + L_{\phi}+L_F,
%\label{eq1}
$$ %\end{equation}
of the  boson $L_B$,  of the quark  $L_Q$ and of the lepton $L_L$ Lagrangians
 \cite{R-99}--\cite{PS-95}.
As far as  $L_B$ and $L_Q$ are not changed we  concentrate our attention on
the lepton  Lagrangian, which for the first lepton generation is written in the form
\begin{equation} 
L_{L,e}=L_l^{\dagger}i\tilde{\tau}_{\mu}D_{\mu}L_l + e_r^{\dagger}i\tau_{\mu}D_{\mu}e_r -
h_e[e_r^{\dagger}(\phi^{\dagger}L_l) +(L_l^{\dagger}\phi)e_r],
\label{eq14}
\end{equation}
where 
$
L_l= \left(
\begin{array}{c}
	\nu_l\\
	e_{l}
\end{array} \right)
$
is the $SU(2)$-doublet,  $e_r $ is the $SU(2)$-singlet, $h_e$ is constant,  
$\tau_{0}=\tilde{\tau_0}={\bf 1},$ $\tilde{\tau_k}=-\tau_k $, $\tau_{\mu}$ are Pauli matricies, 
$\phi \in C_2$ are matter fields and
$e_r, e_l, \nu_l $ are two component Lorentzian spinors.
The covariant derivatives of the lepton fields are given by
$$
D_{\mu}e_r = \partial_\mu e_r +ig'A_\mu e_r \cos \theta_w -ig'Z_\mu e_r \sin \theta_w,
$$
$$ %\begin{equation}
D_\mu L_l=\partial_\mu L_l -i\frac{g}{\sqrt{2}}\left(W_{\mu}^{+}T_{+} + W_{\mu}^{-}T_{-} \right)L_l  -i\frac{g}{\cos \theta_w}Z_\mu\left( T_3 -Q\sin^2 \theta_w  \right)L_l -ieA_\mu Q L_l,
%\label{eq4-2}
$$ %\end{equation}
where $T_k=\frac{1}{2}\tau_k, k=1,2,3$ are generators of   $SU(2)$,
$T_{\pm}=T_1\pm iT_2 $,   
 %$Y=\frac{1}{2}{\bf 1}$ is generator of   $U(1)$, 
 $Q=Y+T_3$, % is generator of electromagnetic subgroup,
 $Y=-\frac{1}{2}{\bf 1}$  is hypercharge of the left leptons,    
$ e=gg'(g^2+g'^2)^{-\frac{1}{2}} \;$ is electron charge and
$ \sin \theta_w=eg^{-1}.$  
The gauge fields  
$$ 
  {W_\mu^{\pm}=\frac{1}{\sqrt{2}}\left(A_\mu^1\mp iA_\mu^2  \right)}, \quad
   { Z_\mu =\frac{1}{\sqrt{g^2+g'^2}}\left( gA_\mu^3-g'B_\mu \right)},
 $$
 \begin{equation}
 { A_\mu =\frac{1}{\sqrt{g^2+g'^2}}\left( g'A_\mu^3+gB_\mu \right)} 
  \label{eq5-1}
\end{equation} 
are expressed  through the fields 
$$ 
A_\mu (x)=-ig\sum_{k=1}^{3}T_kA^k_\mu (x),\quad B_\mu (x)=-ig'B_\mu (x),
$$
which take their values in the Lie algebras $su(2)$,  $u(1)$, respectively.

Next two lepton generation (muon and muon neutrino,  $\tau$-lepton and $\tau$-neutrino) are introduced in a similar way.
Full lepton Lagrangian is the sum
$$ %\begin{equation}
L_L=L_{L,e}+L_{L,\mu}+L_{L,\tau},
%\label{eq14-f}
$$ %\end{equation}
where each term has the structure (\ref{eq14}) with constants 
$h_e, h_{\mu}, h_{\tau}$, correspondingly.

\section{Limiting case of the lepton sector of electroweak model}

The contracted group $SU(2;j)$ is defined \cite{Gr-10-4} as the transformation group
$$ 
z'(j)=
\left(\begin{array}{c}
jz'_1 \\
z'_2
\end{array} \right)
=\left(\begin{array}{cc}
	\alpha & j\beta   \\
-j\bar{\beta}	 & \bar{\alpha}
\end{array} \right)
\left(\begin{array}{c}
jz_1 \\
z_2
\end{array} \right)
=u(j)z(j), \quad
$$
$$ %\begin{equation}
\det u(j)=|\alpha|^2+j^2|\beta|^2=1, \quad u(j)u^{\dagger}(j)=1.
%\label{g3}
$$ %\end{equation}  
of the space $C_2(j)$, which keep invariant the hermitian form
$$ %\begin{equation}
z^\dagger z(j)=j^2|z_1|^2+|z_2|^2,
%\label{g1}
$$ %\end{equation}
where  $z^\dagger=(j\bar{z_1},\bar{z_2}), $ parameter  $j=1, \iota$, and  $\iota$ is nilpotent unit  $\iota^2=0.$ 
The equivalent and more traditional in physics way of group contraction \cite{IW-53} is to tend contraction parameter to zero $j \rightarrow 0$.
%(\ref{g1})
%

The group generators
$ %\begin{equation}     
  T_1(j)= 
j\frac{1}{2}\tau_1, \; 
T_2(j)= 
j\frac{1}{2}\tau_2, \; 
T_3(j)= 
\frac{1}{2}\tau_3 \; 
%\label{g7}
$%\end{equation} 
are subject of commutation relations 
$$  
[T_1(j),T_2(j)]=-ij^2T_3(j), \quad [T_3(j),T_1(j)]=-iT_2(j), 
$$
$$ %\begin{equation} 
 [T_2(j),T_3(j)]=-iT_1(j)
%\label{g8}
$$ %\end{equation}
and form the Lie algebra  $su(2;j)$.
The actions of the unitary group $U(1)$ and the electromagnetic subgroup $U(1)_{em}$ 
in the   fibered  space $C_2(\iota)$ with the base $\left\{z_2\right\}$ and the fiber $\left\{z_1\right\}$ are given by the same matrices as on the space $C_2$.

In the standard electroweak model the  gauge group $SU(2)\times U(1)$ acts in the boson, lepton and quark sectors, i.e. it is the invariance group
of the boson $L_B$, lepton $L_L$ and quark $L_Q$ Lagrangians. We consider a model where the group $SU(2)\times U(1)$ acts only in the boson and quark sectors, whereas in the lepton sector acts contracted group 
$SU(2;j)\times U(1)$. In other words, boson  and quark Lagrangians remain the same as in the standard model, but lepton Lagrangian  is transformed.

The fibered space $C_2(j)$ of the fundamental representation of $SU(2;j)$ group can be obtained from $C_2$ by substituting $jz_1$ instead of $z_1.$
Substitution $z_1 \rightarrow jz_1$ induces another ones for Lie algebra generators
$T_1 \rightarrow jT_1,\; T_2 \rightarrow jT_2,\;T_3 \rightarrow T_3. $
As far as the gauge fields take their values in Lie algebra, we can substitute gauge fields instead of transformation of generators, namely
\begin{equation}
A_{\mu}^1 \rightarrow jA_{\mu}^1, \;\; A_{\mu}^2 \rightarrow jA_{\mu}^2,\; \;A_{\mu}^3 \rightarrow A_{\mu}^3, \;\;
B_{\mu} \rightarrow B_{\mu}.
\label{g14}
\end{equation}  
For the  gauge fields (\ref{eq5-1})  these substitutions are as follows 
\begin{equation}
W_{\mu}^{\pm} \rightarrow jW_{\mu}^{\pm}, \;\; Z_{\mu} \rightarrow Z_{\mu},\; \;A_{\mu} \rightarrow A_{\mu}.
\label{g15}
\end{equation} 
Let us stress that we can substitute transformation of the generators by the transformation of the gauge fields only in the lepton Lagrangian, which is built with the help of $SU(2;j)\times U(1)$ group. In the boson and quark Lagrangians gauge fields are not changed.
%The matter field $\phi_2$ does not transformed as well as its small part $\chi$.
The field
$
L_l= \left(
\begin{array}{c}
	\nu_l\\
	e_{l}
\end{array} \right)
$
is  $SU(2)$-doublet, so its components are transformed in the similar way as components of vector $z$, namely
\begin{equation}
 	\nu_l \rightarrow j\nu_l, \quad e_{l} \rightarrow e_{l}, \quad e_{r} \rightarrow e_{r}. 
\label{g15-1}
\end{equation} 
The right electron field $e_r $ is  $SU(2)$-singlet and therefore is not transformed.
 
The substitutions (\ref{g15}),(\ref{g15-1}) in  (\ref{eq14}) give rise to 
 the lepton Lagrangian  
$$
L_{L,e}(j)=e_l^{\dagger}i\tilde{\tau}_{\mu}\partial_{\mu}e_l +
e_r^{\dagger}i\tau_{\mu}\partial_{\mu}e_r 
- e e_l^{\dagger}\tilde{\tau}_{\mu}A_{\mu}e_l
+\frac{g\cos 2\theta_w}{2\cos \theta_w}e_l^{\dagger}\tilde{\tau}_{\mu}Z_{\mu}e_l
 +
$$
$$
- g'\cos \theta_w e_r^{\dagger}\tau_{\mu}A_{\mu}e_r
+ g'\sin \theta_w e_r^{\dagger}\tau_{\mu}Z_{\mu}e_r
-m_e[e_r^{\dagger}e_l + e_l^{\dagger} e_r] +
%-h_e[e_r^{\dagger}\phi_2^{\dagger}e_l + e_l^{\dagger}\phi_2 e_r] +
$$
$$
+ j^2\left\{ \nu_l^{\dagger}i\tilde{\tau}_{\mu}\partial_{\mu}\nu_l +
   \frac{g}{2\cos \theta_w} \nu_l^{\dagger}\tilde{\tau}_{\mu}Z_{\mu}\nu_l+
%\right.
%$$
%\begin{equation}
%\left.
+\frac{g}{\sqrt{2}} \left[\nu_l^{\dagger}\tilde{\tau}_{\mu}W_{\mu}^{+}e_l +
%\frac{g}{\sqrt{2}} 
e_l^{\dagger}\tilde{\tau}_{\mu}W_{\mu}^{-}\nu_l \right]
% -h_e[e_r^{\dagger}\phi_1^{\dagger}\nu_l + \nu_l^{\dagger}\phi_1 e_r]
\right\}=
$$
\begin{equation}
 =L_{e,b}+ j^2L_{e,f}.
\label{g15-4}
\end{equation} 
We put   
$
\phi=\phi^{vac}=\left(\begin{array}{c}
	0  \\
	\frac{v}{\sqrt{2}} 
\end{array} \right)
$ 
in (\ref{eq14})  and denote electron mass as   $m_e=h_ev/\sqrt{2}$.
Next lepton generations fields are transformed like (\ref{g15-1})
$$ 
\nu_{\mu,l} \rightarrow j\nu_{\mu,l}, \quad
\nu_{\tau,l} \rightarrow j\nu_{\tau,l}, \quad 
\mu_l \rightarrow \mu_l, \quad \tau_l \rightarrow \tau_l, 
$$
$$ %\begin{equation}
\mu_r \rightarrow \mu_r, \quad \tau_r \rightarrow \tau_r.
%\label{g15-5}
$$ %\end{equation} 
The full lepton Lagrangian is the sum
\begin{equation}
L_L(j)=L_{L,e}(j)+L_{L,\mu}(j)+L_{L,\tau}(j)=L_{L,b}+ j^2L_{L,f},
\label{eq14-4d}
\end{equation}
where each term has the structure   (\ref{g15-4}) with the mass 
$m_q=h_qv/\sqrt{2}, \; q=e,\mu,\tau$.

Let contraction parameter tends to zero $j^2\rightarrow 0$, then the contribution  of electron, muon and tau neutrinos as well as their interactions with others  fields to the Lagrangian (\ref{eq14-4d}) will be vanishingly small in comparison  with electron, muon,  tau-lepton  and gauge boson fields. 

An ideal mathematical constructions are physically realized approximately with some errors. 
When contraction parameter $j$ is small, but different from zero, the full Lagrangian of the model
\begin{equation}
L(j)=L_B + L_Q + L_L(j)=L_r + j^2L_{\nu} 
\label{eq14-F}
\end{equation}
%$L(j)=L_B + L_Q + L_L(j)$  
is splited on two parts: the Lagrangian $L_{\nu}$, which include neutrino fields along with their interactions with gauge and lepton fields and Lagrangian $L_{r}$, which include all other fields.
The neutrino fields part $L_{\nu}$ turn out very small with respect to all other fields $L_{r}$ due to the small contraction parameter $j$. So  Lagrangian (\ref{eq14-F}) describe very rare interaction neutrino fields with matter.

In the mathematical language  the  fields space of the standarad electroweak model is fibered after contraction in such a way that  neutrino fields are in the fiber, whereas all other fields are in the base.
In order to avoid terminological misunderstanding let us stress that we  have in view locally trivial fibering, which is defined by the projection in the field space. This fibering is understood in the context of  semi-Riemannian geometry \cite{P-65}--\cite{Gr-09} and has nothing to do with the principal fiber bundle. The simple and best known example of such fiber space is the nonrelativistic space-time with one dimensional base, which is interpreted as  time, and three dimensional fiber, which is interpreted as proper space. 
It is well known, that in nonrelativistic
physics the time is absolutely, while the space properties can be changed in time.
The space-time of the special relativity is transformed to the nonrelativistic space-time when dimensionfull contraction parameter --- 
velocity of light $c$ --- tends to infinity and dimensionless  parameter
$\frac{v}{c} \rightarrow 0$.

Weak interactions for low energies are characterized by the Fermi constant $G_F$. This constant is determined by experimental measurements and turn out to be very small $G_F=10^{-5}\frac{1}{m_p^2}=1,17\cdot 10^{-5}\; GeV^{-2}$. 
Fermi constant is expressed by the parameters of the standard elecroweak model as follows
$$%\begin{equation}
\frac{G_F}{\sqrt{2}}=\frac{g^2}{2m^2_{W}}.
%\label{eq14-9d}
$$%\end{equation}
Since  $m^2_{W}=g^2v^2/4 $, one obtain $v\approx 246\; GeV$ \cite{O-05}.
This "`large"' dimensionfull parameter  enters in Lagrangian of electroweak model 
through the mass terms $m_q=h_qv/\sqrt{2}$ of quarks, electron, muon, $\tau$-lepton, where it is multipied on the free parameters $h_q$. Therefore a priori we can not say are these terms "`small"' or "`large"'.

On the contrary, in the full  Lagrangian $L(j)$ (\ref{eq14-F}) different order terms are appeared due to the small contraction parameter $j^2$ therefore  neutrino fields  and their interactions are small with respect to all other  fields. % and their interactions.   
Probability amplitude for  weak  current interactions which include two neutrinos is multiplied by $j^2$ when $SU(2)$ group is replaced by $SU(2;j)$. 
Therefore Fermi constant, which is the factor of such  amplitude, is just 
the dimensionfull limit %contraction 
parameter of the model, which describe the rarely interacting neutrinos   at low energies.
%The transformation property of the Fermi constant is easily deduced in the form $G_F \rightarrow j^2G_F$. 
If one introduce  the  dimensionfull constant $G_0=\frac{1}{m_p^2}$, then one can approximate  dimensionless contraction parameter $j^2\approx 10^{-5}$.

It is well known that interaction cross-section for neutrinos %with nucleons
increase with energy \cite{O-05},\cite{PS-95}. For energies greater than
$1\, GeV$ this  dependence is linear. The interaction cross-sections are proportional to $G_F^2$, i.e. to $j^4$ for dimensionless parameter. This   leads to the physical interpretation of the contraction procedure as the  decreasing of the interaction cross-section for neutrinos with nucleons when  energy decrease. The dimensionfull limit %contraction 
parameter $j^2G_0$ is interpreted as energy square in that case.

\section{Conclusion}

We explain the  rarely  neutrinos-matter interaction with the help of contraction the gauge group of the lepton sector of the standard electroweak model, leaving the invariance group of the boson and quark sectors untouched. The mathematical contraction procedure 
%for zero tending  parameter 
is connected with the energy dependence of the interaction cross-section for neutrinos. The dimensionfull limit %contraction 
parameter $j\sqrt{G_0}$ is physically interpreted as neutrino energy and the dimensionless contraction parameter at low energies is approximated as $j\approx 10^{-3}$.

 Limiting case of boson sector of the electroweak model for contracted gauge group was discussed in \cite{Gr-10-1}.
The electroweak model based on the gauge group $SU(2;j)\times U(1)$ in boson and lepton sectors have been considered in \cite{Gr-10}. The fiber of this model consist  of neutrino fields and gauge $W^{\pm}$-boson fields.
If the quark sector is also invariant with respect $SU(2;j)\times U(1)$ group, then the $u$-, $c$- and $t$-quark fields are in the fiber too. The definite choice between these versions of the electroweak model can be done in the presence of detailed information on interactions of elementary fields. The preliminary version of the suggested model was discussed in \cite{Gr-10-2}.

This work is supported by the program "`Fundamental problems of nonlinear dynamics"' of Russian Academy of Sciences.

%\newpage

%\section*{References}

\end{document}